\documentclass[conference]{IEEEtran}
\IEEEoverridecommandlockouts
\usepackage{amsmath,amssymb,amsfonts}
\usepackage{algorithmic}
\usepackage{graphicx}
\usepackage{textcomp}
\usepackage{xcolor}
\DeclareMathOperator*{\argmax}{arg\,max}
\usepackage{bbm}

\def\BibTeX{{\rm B\kern-.05em{\sc i\kern-.025em b}\kern-.08em
    T\kern-.1667em\lower.7ex\hbox{E}\kern-.125emX}}
    
\begin{document}


\title{On the Localization of Unmanned Aerial Vehicles with Cellular Networks}

\author{
	\IEEEauthorblockN{Irshad~A.~Meer,~Mustafa~Ozger~and~Cicek~Cavdar}
	\IEEEauthorblockA{Division of Communication Systems, KTH Royal Institute of Technology, Stockholm, Sweden
		\\ \{iameer, ozger, cavdar\}@kth.se}
	
}

\maketitle

\maketitle

\begin{abstract}
Localization plays a key role for safe operation of UAVs enabling beyond visual line of sight applications. Compared to GPS based localization, cellular networks can reduce the positioning error and cost since cellular connectivity is becoming a prominent solution as a communication system for UAVs. As a first step towards localization, UAV needs to receive sufficient number of localization signals each having a signal to interference plus noise ratio (SINR) greater than a threshold. On the other hand, three-dimensional mobility of UAVs, altitude dependent channel characteristics between base stations (BSs) and UAVs, its line of sight and non-line of sight conditions, and resulting interference from the neighboring BSs pose challenges to receive usable signals from the required number of BSs.  In this paper, we utilize a tractable approach to calculate localizability probability, which is defined as the probability of successfully receiving usable signals from at least a certain number of BSs. Localizability has an impact on overall localization performance regardless of the localization technique to be used. In our simulation study, we investigate the relation between the localizability probability with respect to the number of participating BSs, post-processing SINR requirement, air-to-ground channel characteristics, and network coordination, which are shown to be the most important factors for the localizability performance of UAVs. We observe the localizability performance is better at higher altitudes which indicates that localizability with cellular networks for UAVs is more favorable than for terrestrial users.
\end{abstract}

\begin{IEEEkeywords}
localization, unmanned aerial vehicles, cellular networks, interference, air-to-ground channel.
\end{IEEEkeywords}

\section{Introduction}
Utilization of unmanned aerial vehicles (UAVs), also known as drones, has become popular in different non-military and commercial applications such as cargo transport, surveillance and precision agriculture. However, their use especially in the applications requiring beyond visual line of sight (BVLOS) often demands real-time location information of UAVs for their navigation and safe operation  \cite{Ozger_bvlos, meer2019ground}. A widely adopted solution is to use Global Positioning System (GPS) based localization. Nevertheless, its accuracy performance limits its use on UAVs due to the 3-dimensional (3D) mobility of UAVs and highly dynamic environment where they encounter many obstacles during their operation. Furthermore, due to possible large deployment of UAVs, the use of GPS may become costly undermining its potential use. 

One of the connectivity alternatives is the utilization of cellular networks for UAVs, which are termed as \emph{cellular-connected UAVs}. To this end, their integration into cellular networks has become an important research issue \cite{b1, b7,azari2020machine}. Cellular-connected UAVs are considered as aerial users and they coexist with terrestrial users with a certain level of quality of service \cite{b7, azari2017coexistence}.  Consequently, they transmit their application data and have a certain level of reliability for command and control signaling. Due to the mentioned disadvantages of GPS based localization, the cellular connectivity of UAVs can be exploited as an alternative to localize them in the sky. 

Localization via cellular networks has received great attention over the past decades for terrestrial users \cite{b2}. Various localization methods have been devised in 1G to 5G cellular mobile technologies \cite{b3}. These localization techniques utilizing cellular infrastructure based on uplink and downlink communications have different performance in terms of positioning accuracy and required signaling. 

Overall localization performance based on downlink cellular communication depends on number of participating BSs  \cite{b80}. Hence, the first step of the localization process is to make sure that the device can successfully receive localization signals from a sufficient number of BSs. Different localization techniques based on different metrics such as received signal strength indicator (RSSI), time difference of arrival (TDOA), angle of arrival (AOA) and observed time difference of arrival (OTDOA) can be used to compute the position of UAVs. These techniques consider different radio signal measurements or references for the localization process. 

Range based localization techniques using different metrics such as RSSI, TDOA, AOA and OTDOA implement range combining  methods, e.g., trilateration,  triangulation,  or  multilateration. In order for any kind of these localization technique to work, it is important that the UAV receives signals from multiple sources with a signal to interference plus noise ratio (SINR) greater than a specific threshold. In case of the timing advance based localization for UAVs, the estimated time difference translates into the circle around the BS, which suggests the distance between the UAV and the BS. To localize the  UAV, at least three BSs are required for intersection of such circles and for better accuracy more BSs should participate in the multilateration procedure. In case of AOA, the minimum requirement is of two BSs and for TDOA, we require four participating BSs \cite{b83}. It is also established in the literature that there is a relationship between the number of BSs participating in the localization and the system operator's ability to meet the localization performance requirement \cite{b81, b82}. For example, OTDOA based localization system using frequency reuse, require to receive localization signals from  at-least six BSs to meet the requirement of the FCC E911 mandate \cite{b83}. Therefore, each technique has a minimum requirement of the participating BSs. The performance of the localization technique increases with the increasing number of participating BSs. Regardless of the technique, generally localization performance depends on three factors \cite{b80, b60}: \begin{itemize}
	\item relative location of the surrounding base stations (BSs) to the target device,
	\item number of participating BSs,
	\item accuracy of the location observations. 
\end{itemize} To study the performance of localization techniques, the Cramer-Rao Lower Bound (CRLB) can be used in deterministic scenarios \cite{b77}. The CRLB provides a lower bound on the variance achievable by any unbiased location estimator. Although deterministic conditions on the performance evaluation can provide some insights, due to the dynamic nature of wireless channel between users and BSs, generalized views are hard to obtain from it. Hence, only CRLB bounds as a performance metric can be insufficient in most of the cases for the localization via cellular networks.

Authors in \cite{b80} investigate the use of cellular networks for localization of terrestrial mobile devices with the help of stochastic geometry. However, they assume infinite number of BSs which is an impractical assumption for the analytical process. In \cite{iot}, authors have developed an analytical model to investigate the positioning performance of the devices using narrow-band Internet-of-Things (IoT) technology. Apparently, the previous studies focus only on the localization of terrestrial users disregarding emerging aerial users such as cellular-connected UAVs. 

For the localization of terrestrial users, the path loss model is utilized for the calculation of the received power and signal to interference plus noise ratio (SINR) without considering shadowing effect \cite{iot}. Also they assume line of sight (LOS) condition between a participating BS and the terrestrial user with constant path loss exponents. However, for the localization of UAVs, having an LOS condition has a certain probability which depends on the altitude of the UAVs. Furthermore, path loss exponent changes with the altitude for aerial users. Hence, we have a more dynamic channel characteristic. Furthermore, network load highly affects the participation of BSs in localizability.  Hence, there is a strong relation between localization performance and network and wireless channel specific parameters such as the probability of LOS condition between the participating BS and the UAV, altitude of the UAV and network load.     


We utilize the term $B$-localizability as the probability that at least $B$ number of BSs in the network can successfully participate in the localization process \cite{b80}. The minimum number of BSs depend on the employed localization technique. For instance, in case of TOA based technique, we need at least three localization signals from different BSs to locate a target device. Although the localizability performance for terrestrial users has been studied in \cite{b80} and \cite{iot}, it has not been investigated for cellular-connected UAVs by capturing the inherent nature of A2G channels and network dynamics such as interference.    
 
In this paper, with the help of the 3GPP model \cite{b7} for cellular-connected UAVs, we study the localizability performance of a UAV in terms of different network parameters in an urban macro cell scenario. We investigate the impact of the altitude of the UAV on the localizability performance under different conditions such as the number of participating BSs. The effect of network coordination among the BSs on the localizability probability has been investigated. Processing gain, which is an increase in the received SINR by integrating incoming localization signals in time \cite{b80}, is also studied to achieve a certain localizability performance. In our analysis, we utilize the methods for calculating localizability probability proposed in \cite{b80} and  \cite{iot}.  

The paper is organized as follows. Section \ref{sec2} provides information about our system model. Section \ref{sec3} gives the theoretical analysis of the localization problem in terms of localizability. Afterwards, Section \ref{sec4} presents performance results of the localizability probability for different network parameters and channel conditions. Finally, Section \ref{conc} concludes our paper.

\begin{figure}[t!]
	\centerline{\includegraphics[width=\columnwidth,height=4.5cm]{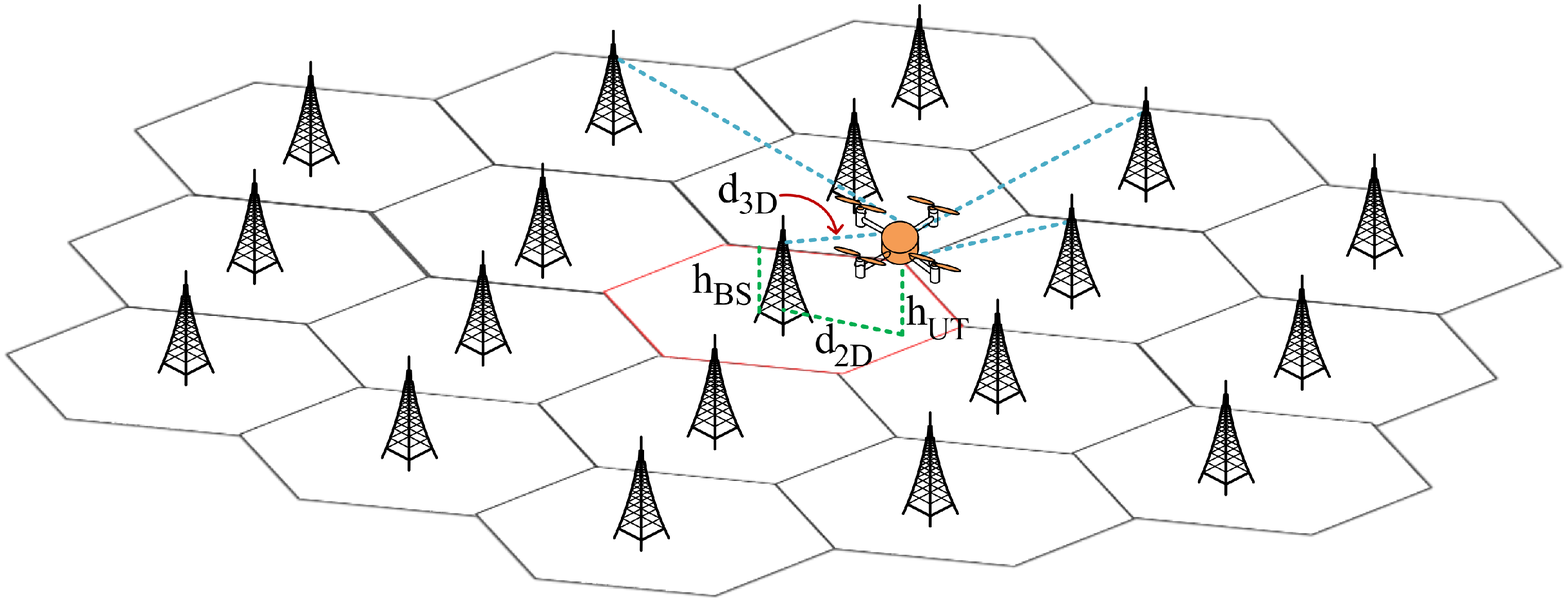}}
	\caption{Two-tier cellular network with hexagonal tessellation, localization signals and distance relations.}
	\label{hex_grid}
\end{figure}

\section{System Model}
\label{sec2}

In this section, the system model for cellular network assisted UAV localization with a hexagonal tessellation is presented. Key notations to explain the system model is provided in Table \ref{table:notation}. We consider a finite two-tier cellular network with $T$ BSs as shown in Figure \ref{hex_grid}, where $T = 19$ in our network. The UAV to be localized is placed uniformly in the central cell at an altitude of $h_{UT}$.  The central cell as marked with red in Figure \ref{hex_grid} is surrounded by ($T-1$) cells. A downlink localization technique such as OTDOA is adopted where a UAV receives the signals from the BSs for the localization process. For the localization of the UAVs, it has to successfully receive at least a certain number of localization signals from the surrounding BSs. In Figure \ref{hex_grid}, the UAV receives localization signals from four BSs as indicated by blue lines for the localization. The green lines show the distance relations in the network.

Since localization signals are short duration signals, we assume that modulation and coding scheme remains same. Interference from the other BSs would then act as the noise and hamper successful reception of the signals. Hence, to investigate the channel quality between the UAV and BS, SINR is the most appropriate metric to study the successful reception of the localization signals. SINR at the UAV has to be greater than a specific SINR threshold to successfully utilize the localization signal. 

The link between a UAV and a BS can have either LOS or non-line of sight (NLOS) condition. Due to the shadowing from the obstacles, the path loss in NLOS link will be higher than the LOS link.  As the altitude of the UAV increases above the ground, the probability of having LOS condition with neighboring BSs increases, and thus better reception of both the useful signal and interference. The probability of having LOS condition with a BS depends on the size and density of the blockage in the environment such as buildings. For our analysis in this paper, we considered a scenario with urban macro cells with aerial vehicles (UMa-AV) characterized in \cite{b7}.  

For the reverse case such that the UAV behaves as a BS, the probability of LOS with the terrestrial users has been defined for altitudes higher than $120$ m \cite{b8}. However, in this work, we adopt the 3GPP model proposed in \cite{b7} for cellular-connected UAVs flying below $120$ m. For the UMa-AV scenario, the probability of LOS, $P^{LOS} $, is given as

\begin{equation}
\begin{aligned}
\displaystyle
	P^{LOS} = \left\{\begin{array}{*{10}{c}}
{1,}&{{d_{2D}} \le {d_1}}\\
{\frac{d_1}{d_{2D}} + \exp \left( {\frac{ - {d_{2D}}}{p_1}} \right)\left( {1 - \frac{d_1}{d_{2D}}} \right),} & {{d_{2D}} > {d_1}}
\end{array} \right.,
\end{aligned}
\label{eq:p_los}
\end{equation} 
where
\begin{equation} {p_1} = 4300\,{\kern 1pt} {\log _{10}}({h_{UT}}) - 3800, \end{equation}
\begin{equation} {d_1} = \max \left( {460\,{\kern 1pt} {{\log }_{10}}({h_{UT}}) - 700,18} \right), \end{equation} $d_{2D}$ is the distance between the BS and the location of the UAV projected onto ground plane, and $h_{UT}$ is the altitude of the UAV as seen in Figure \ref{hex_grid}. $h_{UT}$ can be greater or smaller than the height of the BS, $h_{BS}$. The path loss $L_{m}$, where $m\in[\textrm{LOS}, \textrm{NLOS}]$ for the LOS and NLOS link conditions, respectively,  can be modeled as \cite{b7}:
\begin{equation}
\begin{aligned}
L_{LOS} = 28.0 + 22{\log _{10}}({d_{3D}}) + 20{\log _{10}}({f_c}),
\end{aligned}
\label{eq:l_los}
\end{equation}
\begin{equation}
\begin{aligned}
L_{NLOS} = &  - 17.5 + (46 - 7{{\log }_{10}}(h_{UT})){\log _{10}}(d_{3D})\\
      & + 20{\log _{10}}( {\frac{{40\pi {f_c}}}{3}}). 
\end{aligned}
\label{eq:l_nlos}
\end{equation}

\begin{table}[t!]
\caption{Key Notations Used} 
\centering 
\resizebox{\columnwidth}{!}{%
\begin{tabular}{ l   l  } 
\hline\hline 
Notation & Description  \\ [0.5ex] 
\hline 
$h_{UT}$ & Altitude of the UAV from the ground \\ 
$h_{BS}$ & Height of the base stations  \\
$d_{2D}$  & 2D distance between the UAV and BS    \\
$d_{3D}$   & 3D distance between the UAV and BS \\
$\Phi_i$  & Transmitted power of the BS $i$\\ 
$\zeta$  & Independent shadowing effect \\ 
$\sigma^2$  & Variance of the additive white Gaussian noise \\
$L$   & Path loss between the BS and UAV \\
$T$  & Total number of BSs in the network \\ 
$\mathcal{T}$  & Set of BSs in the network \\ 
$B$  & Number of BSs taking part in the localization of the UAVs \\ 
$W$  & Communication bandwidth \\ 
$f_{c}$  & Carrier frequency used \\ 
${ \mathcal{I}_1}$   & Interference from BSs participating in localization \\ 
${ \mathcal{I}_2}$  & Interference from BSs not taking part in localization  \\ 
${ \mathcal{I}}$ &  Total cumulative interference to the localization of the UAV \\
$\alpha$  & SINR threshold before the processing gain    \\ 
$\beta$  & SINR threshold after the processing gain   \\ 
$\gamma$  & Processing gain required\\ 
$p,q$ &  Activity factor modeling the coordination and network traffic \\
$r_{k},s_{j}$   & Indicator variables \\[1ex] 
\hline 
\end{tabular}
}
\label{table:notation} 
\end{table}

Based on the above channel models, the SINR at the UAV from an $i$th ($i \in \mathcal{T}$) BS  which is at a 3D distance of $d_{i}$ and altitude of $h_{UT}$ is calculated as
\begin{equation}
\begin{aligned}
SINR_{i} = \frac{\Phi_{i}\zeta_{i}L_{m}^{-1}(d_{i})}{\mathcal{I} + \sigma_{i}^2},
\end{aligned}
\label{eq:sinr}
\end{equation} 
where $\Phi_{i}$ is the transmitted power from the  $i$th BS to the UAV, $\zeta_{i}$ denotes the independent shadowing affecting the signal strength from the  $i$th BS to the UAV, and $\mathcal{I}$ is the cumulative interference from the concurrently transmitting BSs excluding the $i$th BS and is calculated as
\begin{equation}
\begin{aligned}
\mathcal{I} = \sum_{k \in\mathcal{T}\: and\: k \neq i} P_{k}\zeta_{k}L_{m}^{-1}(d_{k}),
\end{aligned}
\label{eq:interference}
\end{equation} 
where $d_{k}$ is the distance between the UAV and the $k$th BS ($k \in\mathcal{T}\; \mathrm{and} \; k \neq i$), which are transmitting at the same time. Among the $T$ BSs, we choose $B$ number of BSs ($B \leq T$) based on the strongest average received signal strength to participate in the localization process. However, their successful participation will depend if they have the SINR greater than a threshold. 

We define $\beta $ as the SINR with the gain provided by integrating the incoming localization signals in time at the UAV, i.e., post processing SINR. $\alpha$ is the received SINR without any processing, i.e., pre-processing SINR, which is given in (\ref{eq:sinr}). The processing gain at the receiver is defined as $\gamma = {\beta}/{\alpha}$. This gain can help satisfy the localizability demands for different localization techniques. Hence, in order to participate in the localization process, signals from these $B$ BSs should have SINRs greater than the pre-processing SINR threshold of $\alpha$. 

Among the $B$ BSs participating in the localization, they try to coordinate and attempt to suppress their own transmission when others are transmitting. Since it is not possible to have perfect coordination between BSs, they will transmit their signal when others are also transmitting with a probability $p$. Each of the remaining ($T - B$) BSs due to the load in the network, is transmitting simultaneously with the probability $q$. To incorporate coordination among the $B$ participating BS, and the traffic demands in the ($T - B$) non-participating BSs, we introduce two independent activity parameters $r_{k}$ and $s_{j}$, respectively. Then, the SINR calculated in (\ref{eq:sinr}) can be represented as
\begin{equation}
\begin{aligned}
SINR_{i}(B) = \frac{\Phi_{i}\zeta_{i}L_{m}^{-1}(d_{i})}{\mathcal{I}_1 + \mathcal{I}_2 + \sigma_{i}^2},
\end{aligned}
\label{eq:sinr_new}
\end{equation} 
where 
\begin{equation}
\begin{aligned}
\mathcal{I}_1 = \sum_{k = 1 \: and\: k \neq i}^{B} r_{k}\Phi_{k}\zeta_{k}L_{m}^{-1}(d_{k}),
\end{aligned}
\label{eq:4_14}
\end{equation} 
and 

\begin{equation}
\begin{aligned}
\mathcal{I}_2 =  \begin{cases}
\displaystyle \sum_{j = 1 + B}^{T} s_{j}\Phi_{j}\zeta_{j}L_{m}^{-1}(d_{j}), & \text{if   B \textless \;T } \\
0, & \text{if B = T }
 \end{cases}
\end{aligned}
\label{eq:4_15}
\end{equation} 
where $r_{k}$ and $s_{j}$ are binary variables which are equal to one with probability $p$ and $q$, respectively, and equal to zero with the probability ($1-p$) and ($1-q$), respectively. Therefore, the $SINR_{i}$ is a function of $B$ due to the different activity parameters associated with participating and non-participating BSs.

\section{Localizability Analysis}
\label{sec3}

In this section, we develop a theoretical framework to analyze the localization performance in terms of the localizability probability of the UAVs with the help of the cellular networks. 

Based on the discussions in the previous section, we can investigate if a minimum number of BSs can participate in the localization process given a network layout, coordination and the traffic. As already mentioned, given the pre-processing SINR threshold, $\alpha$, requirement for the localization process, the number of BSs successfully participating in the localization of the UAVs can be investigated. If we define a random variable $\Psi$ as the maximum number of BSs which are successfully participating in the localization process, given our system model we can calculate $\Psi$ as
\begin{equation}
\begin{aligned}
\Psi =\; \argmax_{B \in \mathcal{T}\; and \; B \le{T}} \;  B \times \prod_{i=1}^{B}  \mathbbm{1}\; (SINR_{i}(B) \geq \alpha),
\end{aligned}
\label{eq:4_16}
\end{equation} 
where $B$ is the number of BSs participating in localization and have the strongest signal at the UAV, SINR is given as in (\ref{eq:sinr_new}). $\mathbbm{1}(\theta)$ is the indicator function which is equal to $1$ if $\theta$ is true and equal to $0$ if $\theta$ is false, therefore $\Psi$ will be equal to $B$ when all the signals from $B$ BSs have the SINR greater than the threshold. 


We define $B$-localizability probability as the probability that  at least $B$ BSs successfully participate in the localization procedure \cite{b80}. $B$-localizability probability is denoted by $P_{B}$, and is defined as the probability that $\Psi$ is greater than or equal to $B$ ($Pr(\Psi \geq B)$):
\begin{equation}
\begin{aligned}
P_{B} =\; Pr(\Psi \geq B) = 1\; -\; F_{\Psi}(B),
\end{aligned}
\label{eq:localizability}
\end{equation} where cumulative distribution function (CDF) of $\Psi$, $F_{\Psi}(B)$, is defined as
\begin{equation}
\begin{aligned}
F_{\Psi}(B) =\; P(\Psi \le{B}) = 1\; -\; P(\Psi \geq B)\\
= 1- P \bigg(\big(\prod_{i=1}^{B}  \mathbbm{1}\; (SINR_{i}(B) \geq \alpha)\big)\; =\; 1 \bigg).
\end{aligned}
\label{eq:4_17}
\end{equation} 

In this paper, we will investigate the $B$-localizability performance for the UAVs under different system parameters. We investigate the effect of altitude of the UAVs, network coordination and channel conditions on the  maximum number of participating BSs with SINR greater than the threshold of $\alpha$ for different applicable scenarios. 

For a given network topology, a UAV is said to be $B$-localizable if at-least $B$ BSs successfully participate in the localization procedure. Thus, $P_{B}$ gives us the coverage probability for the various localization techniques (e.g $P_{4}$ for TDOA). To understand the difference between this performance metric (\ref{eq:localizability}) and the traditional metric like CRLB is that, it does not directly give the accuracy but indirectly is an indicator for accuracy. The performance in terms of the error bound given by CRLB do not consider the non-deterministic conditions like network topology and channel condition but only consider a deterministic network with a perfect channel. SINR  as given in (\ref{eq:sinr_new}) captures the effect of the network topology, network traffic, coordination, interference, and most importantly the wireless channel conditions.

\section{Simulation Results and Discussion}
\label{sec4}

We used Monte Carlo simulation and the snapshot model to analyze the effect of the various communication parameters on the $B$-localizability of UAVs.  

\subsection{Simulation Parameters}

In our simulation study, we considered the 3GPP  channel model for the UAVs \cite{b7} for UMa-AV scenario. The UAV to be localized is placed at random in the central cell at different altitudes. The UAV experiences interference from BSs in the second and third layer of the hexagonal tessellation as seen in Figure \ref{hex_grid}. We assume inter site distances to be $500$ m, and  BS height $h_{BS}$ =  $25$ m  \cite{b7}. The bandwidth considered is $10$ MHz and the carrier frequency, $f_c$ is $2$ GHz. The noise figure for the UAV is taken as $9$ dB. The transmitted powers is $46$ dBm. Variance of the shadowing $\zeta$ is modeled as $4.64\mathrm{exp}(-0.0066 h_{UT})$ and $6$ dB for LOS and NLOS conditions, respectively \cite{b7}. Noise figure is assumed to be $9$ dB. 

\subsection{B-Localizability Performance for Different UAV Altitudes}

We study the effect of the UAV altitude on the probability that at least $B$ BSs are successfully participating in the localization for TDOA technique. This effect is studied for different pre-processing SINR thresholds, $\alpha$.

\begin{figure}[h!]
	\centerline{\includegraphics[width=1\columnwidth]{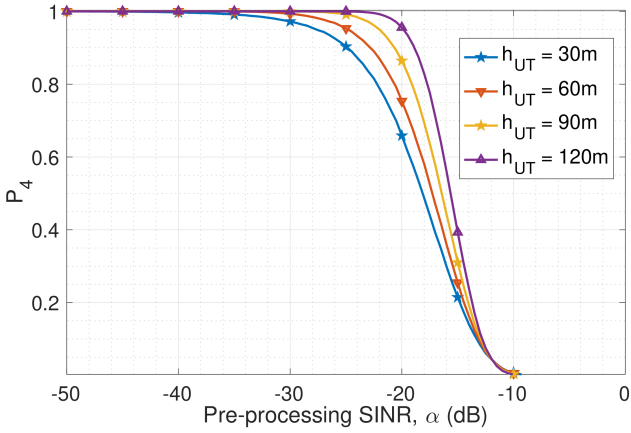}}
	\caption{$P_{4}$ vs. pre-processing SINR threshold, $\alpha$, for the network when $p=1$, $q=1$.}
	\label{fig1}
\end{figure}

Figure \ref{fig1} shows the variation of $P_{B}$ when $B = 4$ with a worst case in terms of network coordination. In this case, all transmissions from participating and non-participating BSs interfere in the localization process, i.e., $p = 1,\; q = 1$. If $\alpha$ is lower than $-40$ dB, $P_4$ is almost one for all altitudes. The reason is that the threshold is so low that the received localization signals by the UAV achieve SINR constraint easily. For SINR thresholds greater than $-10$ dB, the reverse case is observed such that $P_4$ becomes almost zero. On the other hand, for the range of pre-processing SINR threshold between $-40$ dB and $-10$ dB, $P_4$ has lower values for decreasing altitudes. We observe a large difference in the probabilities between the $h_{UT} = 30$ m and $h_{UT} = 120$ m because as we move higher $P^{LOS}$ increases and thus more and more signals can reach the target UAV with sufficient SINR threshold for $-25 \mathrm{~dB} \leq \alpha \leq -15 \mathrm{~dB}$. 

If we consider a processing gain, i.e., $\gamma = 10$ dB,  a post-processing SINR, $\beta$,  is required to be $-6$ dB for successful localization \cite{b60}. Hence, the pre-processing SINR threshold for successfully localizing the UAV is $\alpha = \frac{\beta}{\gamma} = -16$ dB. We see the $B$-localizability probability for $B = 4$ goes from $0.3$ for $h_{UT} = 30$ m to $0.6$ for $h_{UT} = 120$ m when pre-processing SINR threshold $\alpha = -16$ dB. 

\subsection{B-Localizability Performance with Different Number of Participating BSs}

\begin{figure}[!t]
	\centerline{\includegraphics[width=0.9\columnwidth]{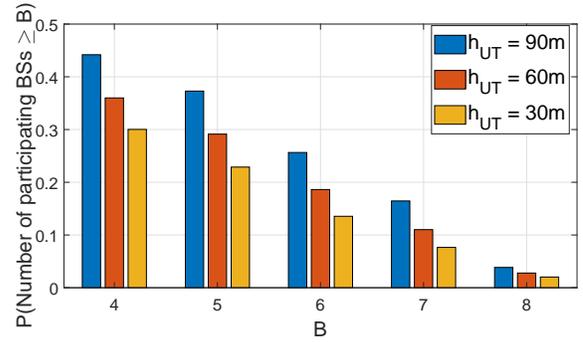}}
	\caption{$P_{B}$ for different $B$ for the network when $p=1$, $q=1$.}
	\label{fig2}
\end{figure}

We analyze the $B$-localizability with change in the number of participating BSs for a pre-processing SINR threshold of $-16$ dB.  Figure \ref{fig2} shows that the  $B$-localizability from $B = 4$ at the UAV altitude $h_{UT}=90$ m is more than $0.4$ in a network that all BSs interfere with the localization signals, i.e., $p = 1,\; q = 1$. We observe that it decreases with a decrease in altitude $h_{UT}$ due to higher path loss experienced in the channel at lower altitudes in NLOS conditions. Furthermore, it indicates that the localizability performance is expected to be poorer for the terrestrial users. Another observation is that as the number of participating BSs, $B$, increases, the $B$-localizability decreases and tends to be zero when $B = 8$. Thus, it is not possible to implement localization techniques which require higher number of participating BSs without any interference mitigation technique. The participating BSs must coordinate to some extent for a better localizability performance and hence localization accuracy.

\begin{figure}[b!]
	\centerline{\includegraphics[width=1\columnwidth]{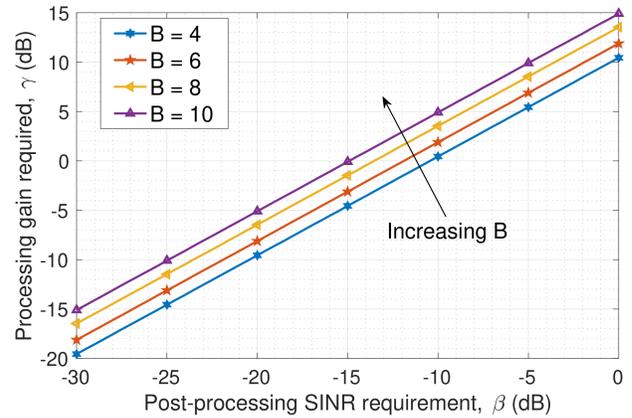}}
	\caption{Processing gain required for achieving $P_{B}=\;0.9$ for different $B$ when $p = 1, q = 1$.}
	\label{fig3}
\end{figure}

\subsection{Processing Gain Requirement}

One way to achieve an acceptable $P_{B}$ is to provide a sufficient gain for the received localization signals.  For a target localizability probability of $P_{B} =\; 0.9 $, we show how the processing gain is changing for a post-processing SINR threshold under different parameters such as altitude and number of participating BSs, $B$. The simulation results are shown in Figure \ref{fig3} for the case of different number of participating BSs. For successful localization of a device, a post-processing SINR requirement of $-6$ dB \cite{b80}. Figure \ref{fig3} shows that a processing gain of $4$ dB to $9$ dB is required for achieving a $P_{B} = 0.9$ with $B$ varying from $4$ to $10$. This gives us the significance of gain provided at the receiver. A major drawback of this is that more gain means more power consumption and more complex circuit components to deploy which may become a challenge for UAVs. 

\begin{figure}[h!]
	\centerline{\includegraphics[width=1\columnwidth]{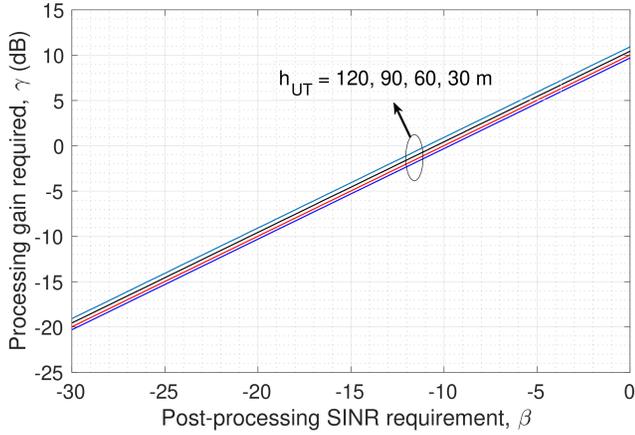}}
	\caption{Processing gain required for achieving $P_{4}=0.9$ for different altitudes when $p=1,~q=1$.}
	\label{fig4}
\end{figure}

Figure \ref{fig4} shows the variation of the gain requirements for different altitudes in the urban macro cell environment for achieving $P_{4} = 0.9$. We see that there is a variation of about $1.5$ dB for UAV altitudes of $30$ m to $120$ m. This shows that for maintaining the same $P_{B}$ with respect to the altitude of the UAV, a small gain is required. Dynamic allocation of gain at the receiver can improve localization performance as well as it can reduce power consumption. Based on the altitude and localization technique, UAV can select the gain for successful participation of required number of BSs. 

\subsection{Localizability Performance with Network Coordination }

In order to demonstrate the effect of interference mitigation through the network coordination among the $B$ participating BSs, we change probability $p$ while keeping the transmission from the non-participating BSs in the worst case, i.e, $q = 1$.

\begin{figure}[ht!]
	\centerline{\includegraphics[width=1\columnwidth]{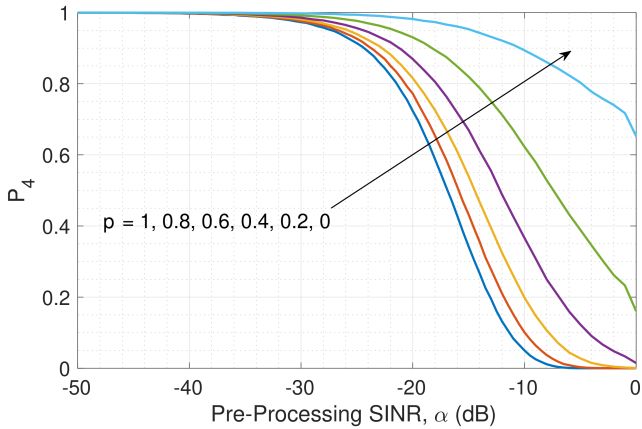}}
	\caption{$P_{4}$ vs. pre-processing SINR threshold, $\alpha$ when $h_{UT} =30$ m, $q = 1$, and $p$ varying from $1$ to $0$ with a step of $0.2$.}
	\label{fig5}
\end{figure}

Figure \ref{fig5} shows that for $B = 4$ and given the pre-processing SINR threshold, $P_{4}$ increases as the coordination among the participating BSs increases. The perfect coordination means that while one is transmitting,  others are not transmitting at the same time. The coordination level increases, i.e., $p$ goes from $1$ to $0$, the probability of $B$-localizability increases. The improvement in the probability of $B$-localizability  indicates the effect of the interference from the surrounding BSs, which can largely be mitigated with network coordination.

\section{Conclusion}
\label{conc}
In this paper, we investigate the $B$-localizability of unmanned aerial vehicles (UAVs), which is the probability of successful reception of localization signals from a $B$ number of base stations (BSs). The $B$-localizability is investigated with respect to UAV-specific parameters such as received interference, air-to-ground channel characteristics including line of sight condition and height dependent channel model, required processing gain at the UAV, number of participating BSs and their coordination to mitigate the interference at the UAV to be localized. This study sheds light on the localization of UAVs in terms of localizability performance with respect to different parameters, which can enable beyond visual line of sight and autonomous operations. Furthermore, we observe $B$-localizability increases with altitude, which can be interpreted that the localizability performance of terrestrial users is worse than that of UAVs. As a future work, we will investigate the effect of mobility on the localizability performance.

\bibliographystyle{IEEEtran}
\bibliography{Ref}

\end{document}